\documentclass[review]{elsarticle}

\usepackage{graphicx,float}
\graphicspath{ {./images/} }
\usepackage{amsmath}
\usepackage{subcaption}
\begin{document}

\author[1]{ Yawer H. Shah\corref{cor1}}
\ead{yawershah@my.unt.edu}
\author[2]{Luigi Palatella}
\author[1]{Korosh Mahmoodi}
\author[3]{Orazio S. Santonocito}
\author[4]{Mariangela Morelli}
\author[4]{Gianmarco Ferri}
\author[4]{Chiara M. Mazzanti}
\author[1]{Paolo Grigolini}
\author[1,5]{Bruce J. West}

\cortext[cor1]{Corresponding author}
\address[1]{Center for Nonlinear Science, University of North Texas, P.O. Box 311427, Denton, Texas 76203-1427}
\address[2]{Liceo Scientifico Statale ``C. De Giorgi"
viale De Pietro, I-73100 Lecce,  Italy}
\address[3]{Division of Neurosurgery, Spedali Riuniti di Livorno, Azienda Sanitaria Toscana Nord-Ovest, 55100 Livorno, Italy}
\address[4]{Fondazione Pisana per la Scienza, Pisa, Italy}
\address[5]{Office of Research and Innovation, North Carolina State University, Raleigh, North Carolina}

\title{Cell Motility in Cancer, Crucial Events, Criticality, and L\'{e}vy Walks}

\begin{abstract}
The analysis of glioblastoma (GB) cell locomotion and its modeling inspired by L\'{e}vy random walks is presented herein. We study such walks occurring on a two-dimensional plane where the walk is similar to the motion of a bird flying with a constant velocity, but with random changes of direction in time. The intelligence of the bird is signalled by the instantaneous changes of flying direction, which become invisible in the time series obtained by projecting the 2D walk either on the $x$-axis or $y$-axis. We establish that the projected 1D time series share the statistical complexity of time series frequently used to monitor physiological processes, shedding light on the role of crucial events (CE-s) in pathophysiology. Such CE-s are signified by abrupt changes of flying direction which are invisible in the 1D physiological time series. We establish a connection between the complex scaling index $\delta$ generated by the CE-s through $\mu_R = 2-\delta$, where $\mu_R$ is the inverse power law index of the probability density function of the time interval between consecutive failures of the process of interest. We argue that the identification of empirical indices along with their theoretical relations afford important measures to control cancer.

\end{abstract}

\begin{highlights}
    
\item We analyze the motility of patient derived glioblastoma (GB) cells using the method of Diffusion Entropy Analysis (DEA). 
\item We find that the trajectories of these cells can be classified as L\'{e}vy walks, with the random changes in direction identified as crucial events (CE-s). 
\item We apply DEA to surrogate time series generated by a two-dimensional L\'{e}vy walk, to prove that the time series formed from one-dimensional projections of the surrogate trajectories yield the correct scaling even though the CE-s are invisible in the 1D projections. 

\item We show that L\'{e}vy walks are a manifestation of criticality generating not only CE-s but also an inverse power law index $\mu_R$, affording information on the time duration of organization,  useful to manage cancer.

\end{highlights}

\maketitle

\section{Introduction} \label{section1}
Glioblastoma (GB) is the most aggressive form of brain cancer and remains the cancer with the bleakest prognosis and the shortest life expectancy (15 months). Less than 6\% of those afflicted survive for five years. A 20 year therapeutic standstill exists for GB, due to the tumor’s ability to develop drug resistance and evade the host immune system demonstrating remarkable ``adaptation to a changing environment”. To this day, no cure for GB is available. The need for new strategies to tackle this tumor is urgently needed. Some relevant emerging properties of intelligent systems are “adaptation to a changing environment,” “reaction to unexpected situations,” “the capacity of problem solving,” and “an ability to communicate.” GB cells are invasive cells that remarkably meet all the above-mentioned requirements. Herein we strip the empirical observations to the bare minimum in order to develop an associated theoretical model.

We begin with the compelling arguments adopted by the researchers of \cite{widelyshared12,collectiveintelligence1,collectiveintelligence2}, and we note that the present paper adopts the same concepts for individual intelligence and collective intelligence as proposed by them to study GB. We share the view of \cite{collectiveintelligence1} on the fact that also the intelligence of a single cell is collective in so far as it is determined by the interactions of the components of the single cells. However, our analysis is done on the cancer cells swimming in a biological fluid and uses the videos of microscopic motion data-set recorded in C. M. Mazzanti's research lab \cite{mazzanti} to analyze the motility and `intelligence" of a single cell while being cognizant of the interdisciplinary nature of this research work \cite{collectiveintelligence2}.

The analysis of dynamics of the single GB cell, as shown using the theoretical approach to criticality, suggests that the single cells themselves have a form of criticality-induced intelligence, or if you prefer a form of \textit{emergent intelligence}. To clarify the concept of emergent intelligence
we quote the results of  a sequence of papers beginning with \cite{vanni}. This paper shows that the transmission of information from one node of a network to all the other nodes of the same network occurs with significant accuracy at criticality that lies between the super-critical and sub-critical conditions thereby affording a significant contribution to the concept of swarm intelligence. The connection between criticality and emergent intelligence is also discussed in \cite{piccinini}. Adopting this key role of criticality, it is possible to address the challenging issue of cognition \cite{grigolini}.

The more recent work of \cite{syncronization} shows that the criticality-induced intelligence generates scaling synchronization between different areas of the same human brain, and among the brain, lungs, and heart of a given individual. Notice that the well-known Dunbar effect
\cite{dunbar1,dunbar2} is also explained to be a consequence of criticality. This construct of intelligence implies an interaction among various units of the same network. In the case of the cancerous cells studied in this paper these units are the biological components of the GB cell, but we do not rule out the possibility that the ``intelligence" of the cell may be influenced by its environment, including its interaction with other cells.
We add to the list of papers contributing to the concept of this form of intelligence the work of \cite{criticalityandintelligence} which focuses on life itself as an emergent property. It is important to notice that 
the concept of such a form of ``intelligence" is interdisciplinary, involving various aspects of social issues \cite{social}.

An important contribution to the concept of emergent intelligence is given in the cancer dynamics context by Wood et al.\cite{widelyshared12} who have found empirical evidence that the pairwise interaction between GB tumor cells generates collective behavior, characterized by scale-free properties, suggesting the occurrence of critical dynamics in their motility and suggesting further the properties a random walk model of cancer cell motility must possess. This is an important contribution to the collective intelligence generated by the interactions among cancerous cells but is distinct from the ``intelligence" envisioned herein since we mainly focus on the ``intelligence" of single cells. 

We note  that \cite{collectiveintelligence1} addresses the crucial issue of the connection between the ``collective intelligence" and the ``intelligence of a single cell", pointing out that the latter form of intelligence emerges from criticality. In this case criticality is due to the interaction among the components of the single cell, a remark of central importance for the research work of the present paper. The discussion on modelling collective behavior in cancer \cite{collectiveintelligence2} is a subject of ongoing and sometimes heated debate involving interdisciplinary perspectives. The Kardar-Parisi-Zhang (KPZ) \cite{renormalizationgroup} is in-and-of itself a phenomenon of organizational complexity \cite{recent,recent2} closely connected to emergent ``intelligence".

In the theoretical approach presented herein criticality generates CE-s time series \cite{crucialevents}. These CE-s time series are interpreted as instantaneous independent corrections of a machine’s malfunction and correcting the malfunction by making the machine brand new with the waiting-time between CE-s being IPL. In this paper the instantaneous correction to the machine's malfunction are quantified by changes of flying directions interpreted as visible CE-s. 

The outline of this paper is as follows. Section \ref{section2} is devoted to modeling the cancer cell motility with L\'{e}vy statistics revealing the intelligence of these cells. Section \ref{section3} details our approach for analyzing the motility data of single cells isolated from patient-derived GB living tissue. In Section \ref{section4}, we interpret the analysis of two-dimensional trajectories as a method of rendering visible the CE-s that, in earlier literature, are often determined to be invisible. Section \ref{section5} addresses the important issue of connecting the CE-s to the regression times so as to establish that the two dimensional trajectories are a manifestation of criticality. These results are confirmed by the numerical analysis of trajectories generated by a well-known form of criticality in Section \ref{section6}. Section \ref{section7} is devoted to summarize the novelty and limitations of the present results leaving open the distinction between the intelligence of a single cell and their collective intelligence.

\section{Connection between L\'{e}vy statistics and CE-s} \label{section2}
The essential features of the model of anomalous diffusion were drawn from the pioneering work of Viswanathan \textit{et~al.} \cite{albatros}, who in 1996 argued that an albatross flying over the ocean in search of food adopts a L\'{e}vy flight pattern. The L\'{e}vy flight strategy is mathematically based on the attractive generalization of the Central Limit Theorem allowing a random walker to make jumps described by an inverse power law (IPL) probability density function (PDF):

\begin{equation} \label{ipl}
\Pi(\xi) = \frac{1}{|\xi|^{\mu}},
\end{equation}
of arbitrarily long jump length $|\xi|$, with the IPL index in the interval $1< \mu<\infty$.

However, our analysis goes beyond the limits of the L\'{e}vy flight approach, originally proposed by Mandelbrot \cite{mandelbrot82}. It is in fact based on  the adoption of the L\'{e}vy walk \cite{0} 
which is compatible with the significant role of CE-s and with the constraint of keeping the scaling smaller than the singular value $\delta =1$.
The adoption of the L\'{e}vy walk approach involves the important property of renewal aging \cite {0} and the singularity of the condition $\mu =2$. The adoption of a L\'{e}vy flight in the case $\mu< 2$ would lead the scaling index to exceed the expected maximum value of 1, while a L\'{e}vy walk would allow us to properly analyse the motility of cancer cells in this case.
The region $1 < \mu < 2$ is characterized by perennial aging and its analysis, based on  observing only the second moment of the diffusion process again yields a violation by producing a scaling index that is $\delta > 1$ \cite{armen}. However, theory using the L\'{e}vy approach makes it possible to establish a connection with the interesting research work of Reynolds et al. \cite{4} whereby they find $\mu = 1.7$ and claim that this is a property of L\'{e}vy walk that correctly represents human mobility. In Section  \ref{section5} we shall establish a closer connection of the view of \cite{4} with the arguments presented herein. 

Despite being compatible with the study of the $\mu < 2$ condition, our statistical analysis of cell motility led us to adopt the condition $2 < \mu <3$, where the CE-s generate a non-stationary correlation function that tends to become stationary in the long-time limit \cite{geisel}, with the form $\frac{1}{t^{\mu -2}}$.

The L\'{e}vy walk we adopt is thus realized by walkers that for an extended time interval $\tau$, called the laminar region, move with constant velocity without changing direction. The time interval between consecutive direction changes is given by the waiting-time distribution density \cite{0}:

\begin{equation} \label{waitingtime}
\psi(\tau) = (\mu-1)\frac{T^{\mu -1}}{(T+\tau)^{\mu}},       
\end{equation}
where, in principle, the IPL index is in the interval \(1 < \mu < 3\). We focus on the interval for the IPL index \(2 < \mu < 3\).

The CE-s occur at the beginning and at the end of the laminar region.  The time series analyzed in earlier works utilizes diffusion entropy analysis (DEA) \cite{idealized} and were obtained by filling the time intervals of length $\tau$, with constant values, 1 and -1, selected with equal probability to realize the 1D random walk. This is a rule close to that adopted in this paper, interpreting the direction  changes  as visible CE-s. In that case, the laminar regions should be filled with constant values denoting the selected flying direction, in accordance with the prescription of \cite{0}. We adopted also different prescriptions that are discussed in subsection \ref{rulesandDEA}.

The survival probability corresponding to this waiting-time PDF has the following analytical form:

\begin{equation} \label{survival}
\Psi(\tau) = \left(\frac{T}{T+ \tau}\right)^{\mu-1}      
\end{equation}
obtained by integrating $\psi(\tau) = - \dot{\Psi}(\tau)$ and inserting the hyperbolic PDF Eq. (\ref{waitingtime})  under the integral for the waiting-time PDF.
Failla et al. \cite{kpz} integrated the time series generated by the process of ballistic deposition yielding the renormalization group phenomenon of Kardar, Parisi and Zhang (KPZ) \cite{renormalizationgroup} to create the diffusion variable $x(t)$ generating the scaling
$\delta$. The scaling $\delta$ is determined by CE-s. The new particles contributing to the growth of the surface fall randomly from the sky. A crucial event occurs when a particle rather than falling on the top of the corresponding column sticks  to the side of one of the next neighbour columns taller that the corresponding column. Why do we call these crucial events? It is a sign that two different columns cooperate to make the surface grow and becomes smooth. In other words, the two columns contribute to the goal of the complex system as the boids of \cite{criticalityandintelligence}, namely the units, not necessarily intelligent, contributing to the swarm intelligence. The authors of \cite{kpz} evaluated the distance between two consecutive CE-s and found for the corresponding waiting time distribution density the IPL index $\mu_S = 1.666$. They found that the scaling $\delta$ is determined by

\begin{equation} \label{SVM}
    \delta = \frac{\mu_S -1}{2}.
\end{equation}

In addition to evaluating the scaling $\delta$ these authors studied also the regression to the origin of the diffusion variable $x(t)$. The time distance between two consecutive regressions to the origin is characterized by a waiting time distribution density with the same inverse power law distribution structure as Eq. (\ref{waitingtime}) with $\mu = \mu_R$,
and rigorous mathematical arguments based on the assumption that the KPZ phenomenon is renewal, yield the following equation:

\begin{equation}  \label{Exact}
\mu_R = 2 - \delta .
\end{equation}

It is important to notice that theoretical results of \cite{kpz} is valid under the condition $\delta < 1$. As a consequence it is not restricted 
to the sub-diffusion condition $\delta < 0.5$. It applies also to the super-diffusion case $\delta > 0.5$, provided that the condition $\delta < 1$ is fulfilled.  

In this paper we argue that $\mu_R$ can be interpreted as affording information on the vulnerability of the complex organization under study 
and we extend the KPZ arguments to the motility of cancerous cells.

\subsection{Walking rules and Diffusion Entropy Analysis} \label{rulesandDEA}
To make it easier for the readers to understand the connection with \cite{0} and the extension of the analysis of dynamical properties described in \cite{0}, we afford here a concise illustration of the tools of statistical analysis that are not explicitly 
mentioned in \cite{0}. 

First of all, we have to illustrate the method of Diffusion Entropy Analysis (DEA). This method was proposed in 2001 in \cite{dea1}.

When the scaling condition 
\begin{equation} \label{globalscaling}
p(x,t) = \frac{1}{t^{\delta}} F \left( \frac{x}{t^{\delta}}\right) ,
\end{equation}
applies, the adoption of the Shannon-Wiener entropy yields:
\begin{equation} \label{shannon}
 S(t) = -\int \mathrm{d}x p(x,t) \mathrm{ln} ( p(x,t) ) = A  + \delta \:\mathrm{ln} t,
 \end{equation}
where A is a constant and the slope of $ S(t) $ vs $ \mathrm{ln} t$ is the scaling $\delta$.
Notice that $x$ is the diffusion variable obtained by integrating over time the experimental time series $\xi(t)$. This method requires the use of a Gibbs ensemble of time series $\xi(t)$. To bypass this limitation the user of DEA also adopts the method of moving window that splits the time series into many distinct parts including in some cases also a small overlap. In this paper we make the assumption that different cells can be interpreted as a Gibbs ensemble and to bypass the limits of not sufficient statistics, we make a joint use of Gibbs ensemble and moving window method. The method of analysis adopted to evaluate the complexity of the cancerous cell is equivalent to fill the laminar regions with either $1$ or $-1$, in accordance to the velocity model of \cite{0}. 

To understand the importance of our discovery of the complexity parameter of cancer cells, it is necessary to make the readers aware of another walking rule, corresponding to assuming that diffusion is generated by making a jump ahead or backward with equal probability. The use of DEA in this case is discussed in \cite{dea2}. The authors of \cite{kpz} adopted this rule to find $\mu_S$ and $\mu_R$. This rule is the Jump model of \cite{0}. The paper \cite{dea2} proposed also the model of making always a jump ahead when a crucial event occurs. The adoption of this rule is important to shed light into the Dunbar effect \cite{dunbar1,dunbar2} and into the vulnerability of cancer cells. It is also important to quote \cite{dea3}, which affords a method to detect crucial events when they are not visible, as in the case of the Dunbar effect \cite{dunbar1, dunbar2}.

\section{Analysis of real video data} \label{section3}

\subsection{Perrin approach to Brownian motion} \label{Perrin}

To facilitate the readers to understand the novelty of our approach we concisely review the research work done by Perrin to empirically confirm Einstein's theory of diffusion.

\begin{figure}[H] 
\centering
\includegraphics[width=0.5\textwidth]{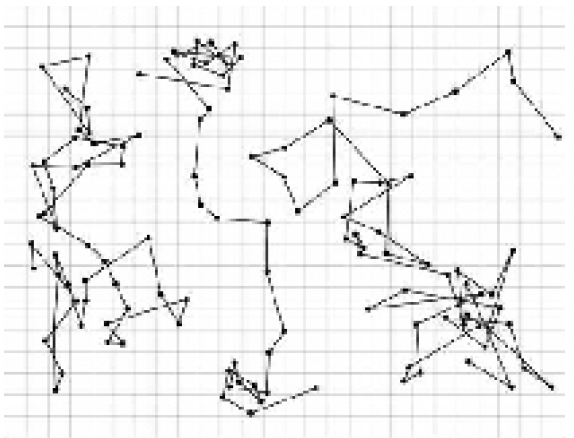}
\caption{The trajectories of a particle undergoing Brownian motion as observed through a microscope by Perrin, where he marked the position of the particle at equal time intervals and subsequently connected the dots by hand. Public domain.}
\label{perrin} 
\end{figure} 

\noindent Figure \ref{perrin} depicts distinct trajectories of Brownian particles being pushed around by thermal excitations of the lighter particles of the ambient fluid. These are the experimental data points that Jean Baptiste Perrin used to empirically establish the validity of Einstein's theory of diffusion and subsequently the existence of atoms \cite{perrin13} and for which he was awarded the 1926 Noble Prize in Physics: ``for his work on the discontinuous structure of matter, and especially for his discovery of sedimentation equilibrium".

\begin{figure}[H] 
\centering
\includegraphics[width=0.5\textwidth]{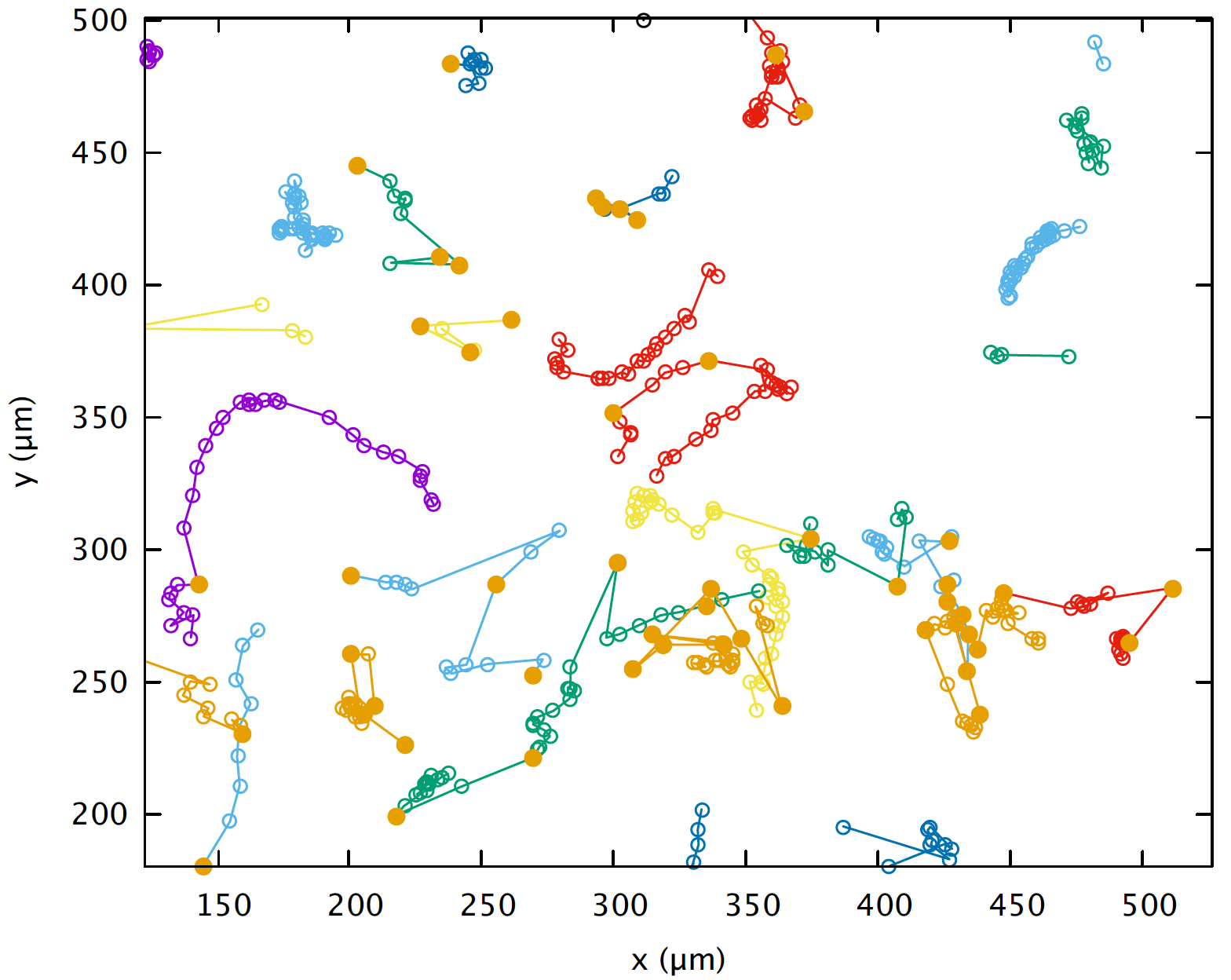}
\caption{The trajectories of cancer cells derived from a patient suffering from GB, with empty circles marking the position of the cancer cells at consecutive times. Each trajectory denotes the track of an individual cell. The filled circles mark the CE-s where the fluctuations in the velocity are larger than a given threshold, $\left| \Delta \vec{v} \right| > 0.013 \mu m$/s.}

\label{fig:2 } 
\end{figure}

In seeking to understand the behavior of microscopic entities one
can do worse than adopt the strategy of Einstein and Perrin, even when the
entities to be described are microscopic living matter swimming in a bio-fluid.
We find that in spite of the wondrous experimental apparatus brought to
bear on the determination of the motility of these microscopic entities that
the trajectories obtained look fundamentally the same as those of Perrin. This forces us to analyze the videos of the lab of C. M. Mazzanti's research lab using our concept of CE. We plot Figure \ref{fig:2 } making the assumption that the change of flying direction are CE-s. For that purpose we mark with a filled circle the points of the two-dimensional trajectories where the velocity change exceeds a given threshold, $\left| \Delta \vec{v} \right| > 0.013 \mu m$/s. We are back to the concept of single cell and possibly collective intelligence. In the next Section we prove that these are ``visible" crucial events.

\subsection{Observations from Experiment} \label{realdata}

The procedure followed in C. M. Mazzanti's research lab, Laboratory of Genomics and Transcriptomics,  was to dissociate
patient-derived GB living tissues into single-cell suspensions by
combining mechanical dissociation with enzymatic degradation. Single-cell
suspension can subsequently be maintained live under carefully controlled conditions by plating them on the bottom of a 25 $cm^2$ culture flask and cultured at 37°C in a liquid medium composed of nutrients, growth factors, cell specific supplements and antibiotics (96 \% DMEM/F12, 1\% Penicillin/Streptomycin, 1\% Amphotericin B, 1\% G-5 supplement, 1\% Glutamax).
The GB cells are recorded for 18 hours through an optical
microscope equipped with an incubation chamber. A total of 220 frames
(1 frame/5 minutes) are recorded in phase contrast mode, and at least 4
time-lapse videos are captured on different fields of view of the same sample
simultaneously in tiling acquisition mode. Time-Lapse videos are analyzed
with Cellpose software for cellular segmentation and TrackMate (ImageJ
plugin) for tracking analysis. Cell trajectories are automatically identified
with correspondences determined in consecutive images, and cell divisions
detected. Each identified trajectory is manually reviewed and six parameters describing morphology and dynamic traits of imaged cells are obtained
(i.e. circularity, area, mean speed, total distance travelled, persistence and
track displacement, compared to Perrin's single data point every 30 seconds).
The important thing to note here is that this process is concluded within five
days from the date of surgery to avoid tissue divergence from the original tumor. This approach enables quantifying important biological processes such
as cell migration and cell growth, both of which play a major role in the
development of diseases.

\subsection{Statistical analysis of the dynamics of cancer cells}
The real data of Figure 2 are a projection onto a 2D plane of a 3D
motion within a fluid medium. We ignore the effect of limited fluctuations
along the direction defined by gravity and evaluate the instantaneous velocity
intensity. When the fluctuations of these instantaneous velocity values exceed
a fixed threshold the occurrence of this fluctuation is interpreted as a CE corresponding to a change of direction in the cell's motion.

Let us shortly review the issue of scaling evaluation by using the simplified picture of one-dimensional walkers. They start from the origin $x = 0$ at the same time $t=0$. Some walkers may move to the right and others may move to the left. Upon time increase the original distribution of walkers, a delta of Dirac at $x =0$, becomes a broader and broader cloud. The width of this cloud is proportional to $t^{\delta}$, where $\delta$ denotes the scaling index.
For the scaling evaluation to be accurate, we require a large number of runners. In principle it is possible to evaluate the scaling $\delta$ with precision using only one runner (one trajectory), provided that the trajectory is extremely extended in time, using the moving window method. The moving window method is based on splitting the single trajectories in many different windows of the same temporal size. The beginning of the window is interpreted as the time origin $t=0$ and the position of the runner, regardless its value, is interpreted, as the space origin $x = 0$. The difference between the position of the runner at the end of window and the position of the runner at the start of the window is the amount of space travelled by the runner. This makes it possible to create many trajectories that are supposed to be independent although determined by the same walking prescription.  

The data under study are generated by observing the motility of different cells. Therefore, in principle we may have adopted the prescription of evaluating the scaling $\delta$ without involving the mobile window method. Unfortunately in each cell trajectory the number of CE-s (visible changes of direction) is less than one hundred and this is not enough for an accurate scaling evaluation. So to capture the statistics, we use mobile windows generated by different cells.

In conclusion, the coordinates of the \textit{k-th} cell at time $\tau$ in this 2D framework are
$ X_k (\tau)$ and $ Y_k (\tau)$ and using the method of the mobile temporal window, we generate several diffusive trajectories $\Delta X_{k,i}(t)$, $\Delta Y_{k,i}(t)$
at different temporal size of the mobile windows, denoted by the symbol $t$, not to be confused with the absolute time $\tau$.
\begin{equation} \label{Discrete}
\begin{array}{l}
 \Delta X_{k,i}(t) =  X_{k}(t+i) -  X_{k}(i) \\
 \\
 \Delta Y_{k,i}(t) =  Y_{k}(t+i) -  Y_{k}(i),   
\end{array}
\end{equation}
where $ 0 \le i < \tau_{max} - t $ is the index indicating the starting
position of the mobile window that runs throughout the entire sequence.  We make the assumption that the statistics of $\Delta X_{k,i}$ is identical to the statistics
of $\Delta Y_{k,i}$.  This assumption is proved to be correct by the theoretical arguments of Section 4.

We evaluate the total number of runners afforded by our approach as follows.  Supposing  that each cell sequence
is $\tau_{max}$ long, for any temporal window of size $t$  
we have a total number of $2 N_{cells}(\tau_{max} -t)$
trajectories positions, denoted by the symbol $x$. We then compute 
the PDF $ p(x , t)$. Obviously, at
different times $ t$ the number of diffusive trajectories can be different  but this
problem is straightforwardly solved by properly normalizing $ p(x , t)$ 
to the standard condition
\[
\int p(x,t) dx = 1.
\]

In Figure 3, we show the results obtained using DEA as illustrated by Eq. (\ref{shannon}). Note that the scaling $\delta = 0.67$ 
is connected to the occurrence of CE-s of Figure (\ref{fig:2 }) as discussed in the next subsection.

\begin{figure}[H]
\centering
\includegraphics[width=0.5\textwidth]{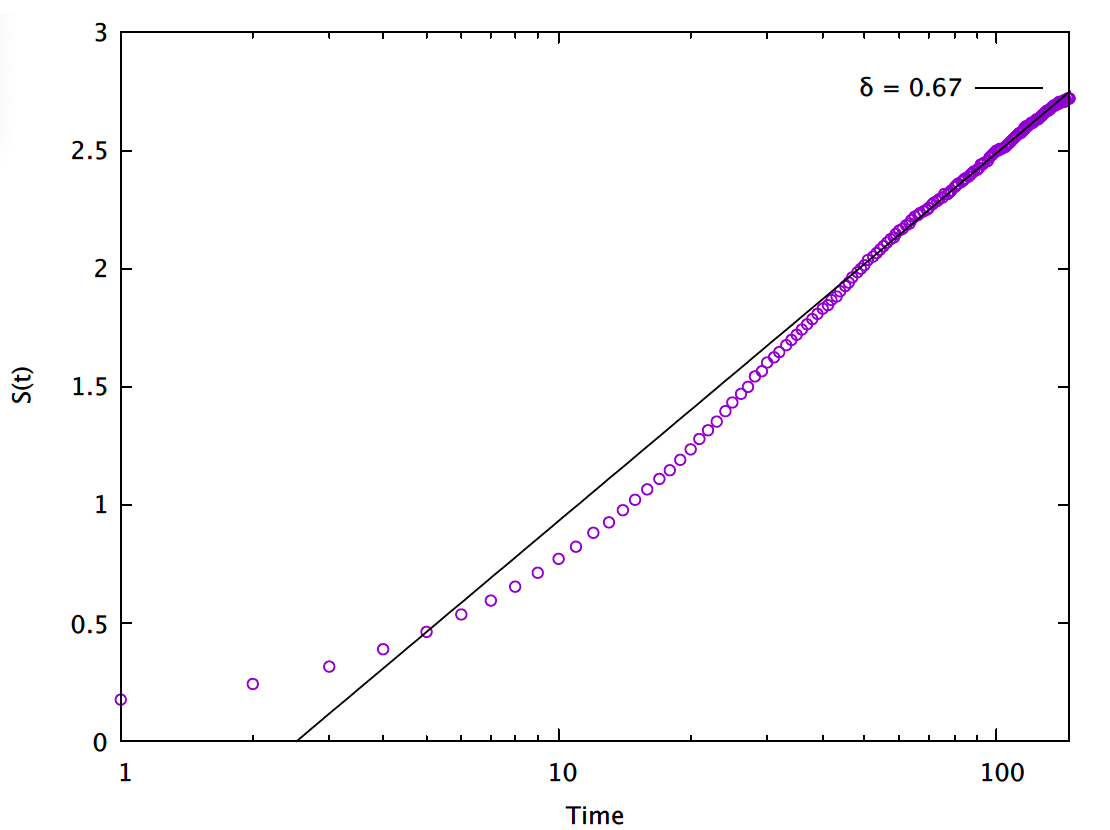}
\caption{We show the results afforded by the technique of Diffusion
Entropy Analysis (DEA) for the trajectories obtained using the procedure of Section \ref{section3} on the cancer cell (Astrocytes) motility data
for a given subject. The straight line segment has a slope fitted to the
asymptotic data points yielding $\delta$ = 0.67 or $\frac{2}{3}$ which using
Eq.(\ref{levy8}) yields $\mu_S$ = 2.5.}
\label{fig:3 }
\end{figure}

\subsection{Remarks on the statistical analysis of experimental results}

The choice of different threshold values will not influence the evaluation of the scaling $\delta = 0.67$, which is obtained by applying the DEA without using stripes for the visible CE-s. The candidates for the state of CE are the visible changes of flight direction, illustrated in Figure \ref{fig:2 }. Note that the L\'{e}vy scaling $\delta$, in accordance with the velocity model of \cite{0} is given by:
\begin{equation} \label{levy8}
\delta = \frac{1}{\mu - 1}.
\end{equation}

\begin{figure}[H]
\centering
\includegraphics[width=0.5\textwidth]{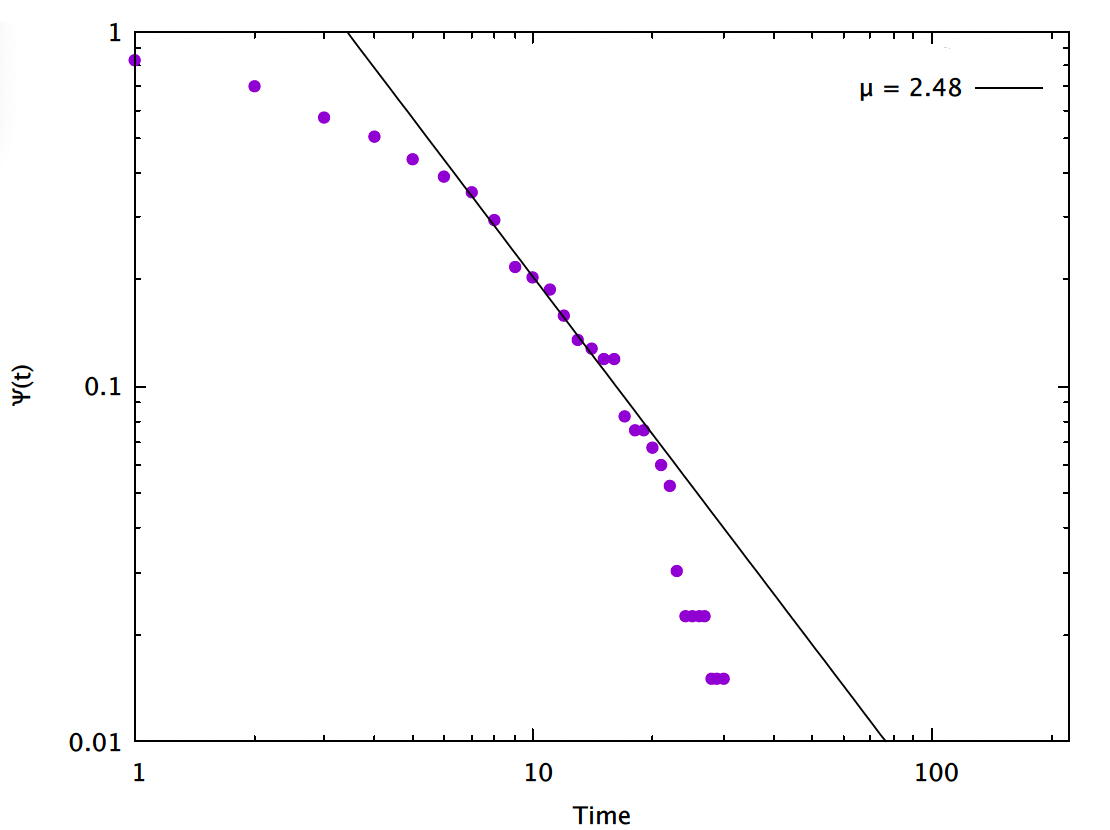}
\caption{The survival probability generated from the cancer cell trajectories measured for the cells of a patient suffering from GB, for events associated with a change in velocity greater than a given threshold, $| \vec{V}(t + \tau) - \vec{V}(t)| >  0.013 $ $\mu m$/s. The straight line segment fitted
to the survival probability gives the IPL index  $\mu$ = 2.48 in good agreement with the results of Figure 3. }
\label{fig:4 }
\end{figure}

To make our conclusion compelling, we have to prove that the distance between two consecutive changes of direction generates the index $\mu = 1 + \frac{1}{\delta} $, that according to the results of Figure \ref{fig:3 } should be $\mu = 2.49$. We know that a renewal process with $\mu < 3$ is characterised by an ergodicity breakdown \cite{0}. However, the time distance between two consecutive CE-s is a property of the brand new waiting-time probability distribution. As a consequence, the inverse power index of the survival probability $\propto \frac{1}{t^{\mu-1}}$ is expected to afford a fairly reliable information on the index $\mu$. We also know that $\Psi(t)$ being a time integration over $\psi(t)$ is computationally more accurate than the direct use of $\psi(t)$. The result of this analysis shown in Figure \ref{fig:4 } leads us to $\mu = 2.48$. We believe that this result is close enough to the value obtained from the DEA method, and this leads us to conclude that the changes of direction are indeed the visible CE-s. We note that the results of the analysis are very close to the ideal value of $\mu = 2.5$. In section \ref{section5} we explain why the value of $\mu = 2.5$ is an ideal value.

Note that we work with integer times. The function $\Psi(\tau)$ is obtained by filling the bins $n= 0, n=2, ....$ with numbers. For any $\tau$ we fill the first $\tau +1$ bins with the number $1$. Each bin $n$ is assigned  the value $N$,which is the number of $\tau 's$ of length equal or larger than $n+1$.  Finally the distribution is normalized by dividing the number assigned to each bin by the value of the bin $n =0$, so as to fulfill the condition $\Psi(0) = 1$.

It is convenient to notice that dynamics between two consecutive filled circles of Fig. 2 are not a straight line  but the fluctuations around the straight lines are of modest intensity so as to qualitatively reproduce the structure of the diffusion trajectory of Fig. 5.

\begin{figure}[H]
\centering
\includegraphics[width=0.5\textwidth]{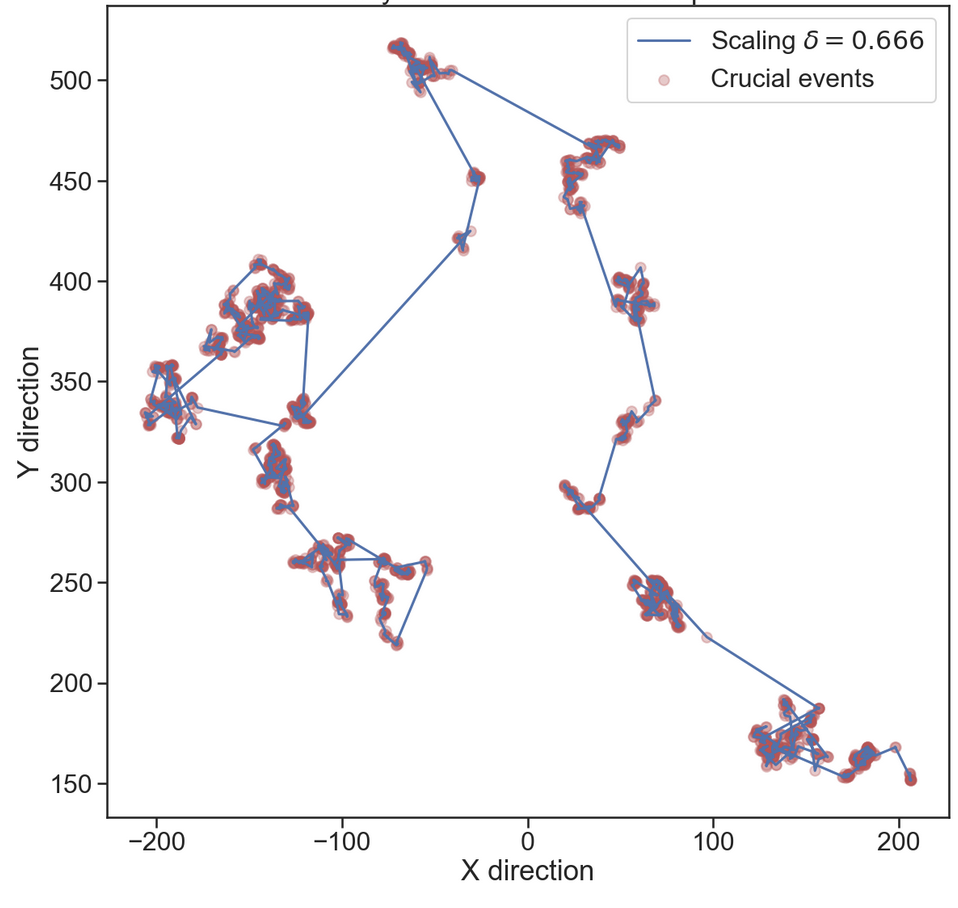}
\caption{We show the 2D Lévy walk generated by the idealized Manneville
map with $\mu$ = 2.5. The motion is in two spatial dimensions with the changes
of direction depicting CE-s, marked here with small circles.}
\label{fig:5 }
\end{figure}

\section{Detection of invisible crucial events} \label{section4}

We devote this section to illustrate the two-dimensional L\'{e}vy walk emerging from the analysis of the data of the research group of C. M. Mazzanti.
In this model the
CE-s are visible as shown in Figure \ref{fig:2 } and Figure \ref{fig:5 }. The CE-s of Figure \ref{fig:5 } are generated using the ideal value of $\mu =2.5$.

Following the suggestions of \cite{bellazzini} we generate a sequence of times $\tau_1, \tau_2,...$  using the idealized Manneville map yielding:

\begin{equation} \label{tau}
    \tau = T \left[ \frac{1}{y^{\frac{1}{\mu -1}}} - 1 \right] ,
\end{equation}
where the number $y$ is drawn with equal probability from the interval [0,1]. The distance between two consecutive changes of direction is $\propto$ $\tau$. In the one dimensional case, each distance is filled with the values $W$ and $-W$ randomly selected with equal probability. The adoption of DEA was proved \cite{bellazzini} to lead to the scaling of Eq (\ref{levy8}).

To generate Fig. 5 we fill the distances with the angle $\theta$ randomly selected using a PDF  uniformly distributed over the interval
$0 < \theta <  2\pi$. In this case, the change of direction is a visible CE, and the visible changes of direction of motion depicted in Figure 5 enable us
to illustrate the dynamical effects of the distances adopted for the
walk. 

Now let us project the trajectory of Fig. 5 onto the $x$ and $y$ axes. This raises a new problem for the DEA analysis. Rather than making the ensemble average adopted by us to get $\delta = 0.67$ we are forced to use the technique of mobile window \cite{dea1, dea2, dea3}. Since, the projections inherit the complexity of the 2D trajectory generated by visible CE-s, we do not need to use the \cite{dea3} to detect CE-s. Hence, we proceed with the traditional method of using DEA without stripes. It is important to notice that despite the aging effect generated by the use of a finite value of $\mu$, the use of mobile windows for $2 < \mu <3$ yields the correct scaling $\delta$ of Eq. (\ref{levy8}). The effect of aging is perceived only for $\mu < 2$ \cite{armen}. We have to notice that the projection of the trajectory of Fig. \ref{fig:5 } does not generate the alternate sign sequence of \cite{bellazzini}, since the projection has the effects of affecting both the length of the laminar regions and the value assigned to the laminar region that is $ \theta $ rather than $\pm W$. 

\begin{figure}[H]
\centering
\includegraphics[width=0.5\textwidth]{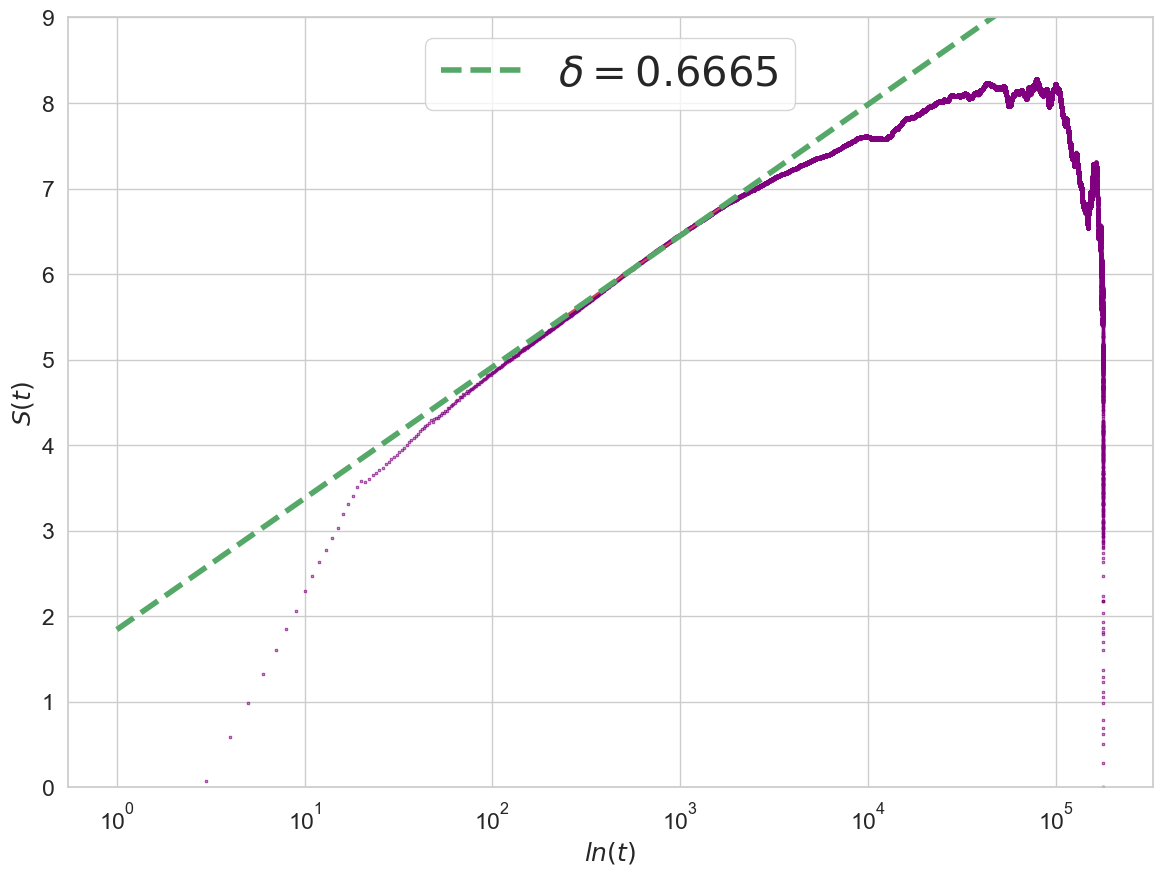}
\caption{The DEA technique (without stripes) is
applied to the projection of the 2D trajectory of Figure \ref{fig:5 }, with visible CE-s, onto the $x$-axis. The time-dependent entropy S(t) versus the
log of the time yields a slope of $\delta$ = 0.6665 represented by the dashed line.
This should be compared with the theoretical results in Figure \ref{fig:8 }. }
\label{fig:7 }
\end{figure}

\begin{figure}[H]
\centering
\includegraphics[width=0.5\textwidth]{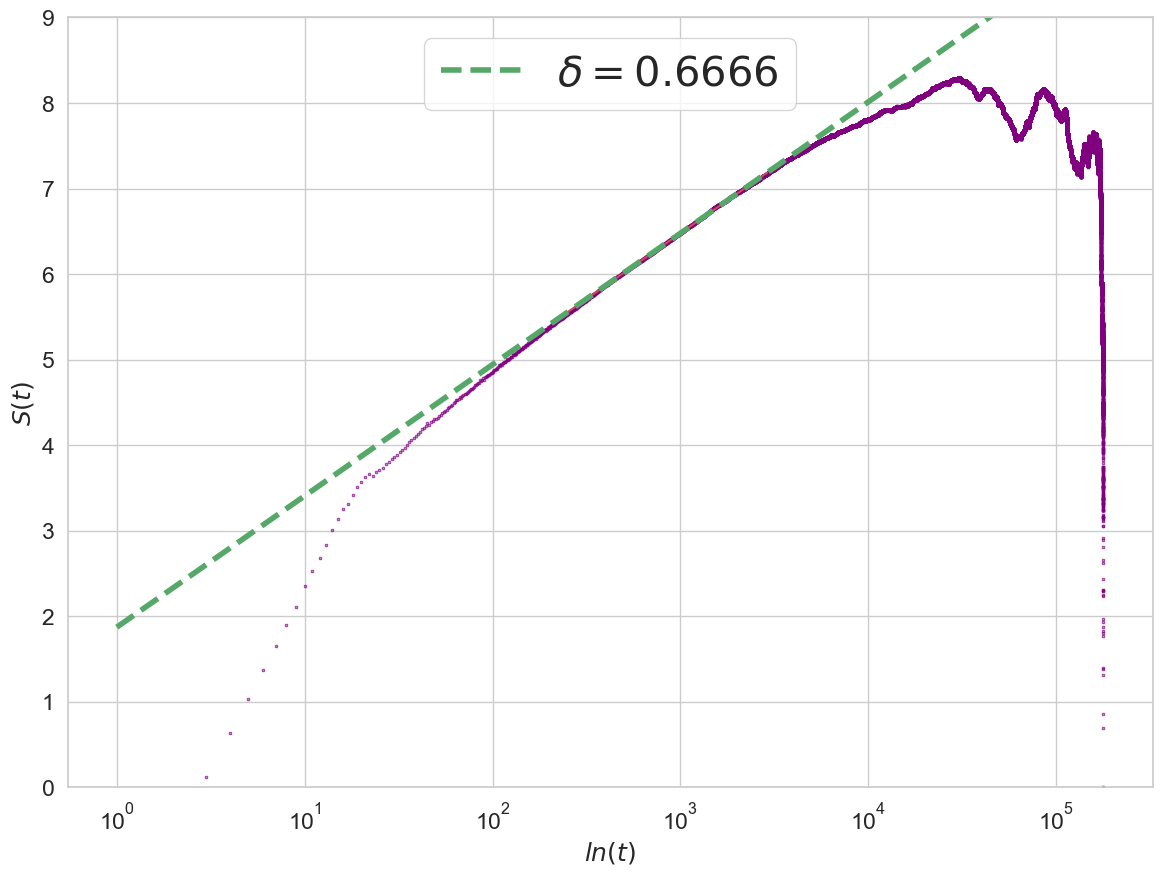}
\caption{The DEA technique (without stripes) is
applied to the projection of the 2D trajectory with visible CE-s onto the $y$ axis. The time-dependent entropy S(t) versus the
log of the time yields a slope of $\delta$ = 0.6666 shown by the dashed line. This should be compared with the theoretical results in Figure \ref{fig:7 }.}
\label{fig:8 }
\end{figure}

Despite these doubts the results of Figure \ref{fig:7 } for the $x$ axis and Figure \ref{fig:8 } for the $y$ axis confirm that the scaling established by DEA is

\begin{equation} \label{levyscaling}
\delta = \frac{1}{\mu_S -1} ,
\end{equation}
with $\mu_S = 2.5$. We use the symbol $\mu_S$ rather than $\mu$ to establish a connection with next Section devoted to discuss $\mu_R$, another important inverse power law index. 

It should be clear that the projection of a trajectory to a lower dimension hides the existence of CE-s but it does not lose any information. If
such a projection did lose information during the projection process it would
need to dynamically interact with the data to erase or add information and
thereby damage or destroy the CE information. It is somewhat surprising to
us however that a passive projection process retains all the statistical information contained in a CE time series. But that is apparently the situation and this paradoxical condition requires a comment. The discovery that the projections of the two-dimensional motion over 
the $x$ axis and $y$ axis generate the identical scaling $\delta = 0.666$ lead us to again quote Reynolds et al. \cite{4}.
To facilitate the identification of movement patterns resembling L\'{e}vy walks, the authors of this article applied a procedure based on the assumption that the projection of a L\'{e}vy walk is itself a L\'{e}vy walk. The results illustrated by Figure \ref{fig:7 } and Figure \ref{fig:8 } confirm this property, thereby reinforcing the claim that DEA serves the useful purpose of revealing the occurrence of CE-s, even when they are not visible. 

\subsection{Origin of truncations}

To support our claim that L\'{e}vy walk of the cancer cell is characterized by the ideal value $\mu_S = 2.5$, corresponding to  $\mu_{R} =1.33$ we must address the intriguing issue of the truncation of the survival probability that is evident in Figure \ref{fig:4 }.
We stress some crucial properties of the influence exerted by the biological fluid on the cancer cells while they swim. The authors of the 1995 paper of \cite{bettin} studied the effect that the environment may have on the anomalous diffusion of a L\'{e}vy walk with $2 < \mu_S < 3$. They found that the waiting time distribution density of Eq. (\ref{waitingtime}) is replaced by
\begin{equation}
\psi(\tau) = (\mu_S - 1) \frac{T^{\mu_S-1}}{(T + \tau)^{\mu_S-1}} e^{- \Gamma \tau},
\end{equation}
with the effect of making in the long time limit the diffusion process normal, although characterized by a very large diffusion coefficient.

The authors of the more recent paper of \cite{pla} expressed the same property using the survival probability $\Psi(t)$ rather than the waiting time distribution density:
\begin{equation} \label{theoreticalsurvival}
\Psi(\tau) = \frac{T^{\mu_S-1}}{(T + \tau)^{\mu_S-1}} e^{- \Gamma \tau},
\end{equation}
and studied the condition $\mu_S < 2$ rather than the condition $2 < \mu_S < 3$ of this paper, thereby missing the observation done in \cite{bettin}  that in this case at times $t > 1/\Gamma $ the normal diffusion is characterized by an anomalously weak diffusion coefficient. 

It is very important to notice that also the work of \cite{2} affords an argument for the truncation of L\'{e}vy walk. The Brownian-like steps self-organize into truncated L\'{e}vy walks to have a functional advantage in random searching. We think that this argument is the most convenient for the motility of cancerous cells analyzed in this paper, since both the adoption of L\'{e}vy walk and its truncation suggest the ``intelligence" of cancerous cells searching for nutrients. 

\begin{figure}[H]
\centering
\includegraphics[width=0.5\textwidth]{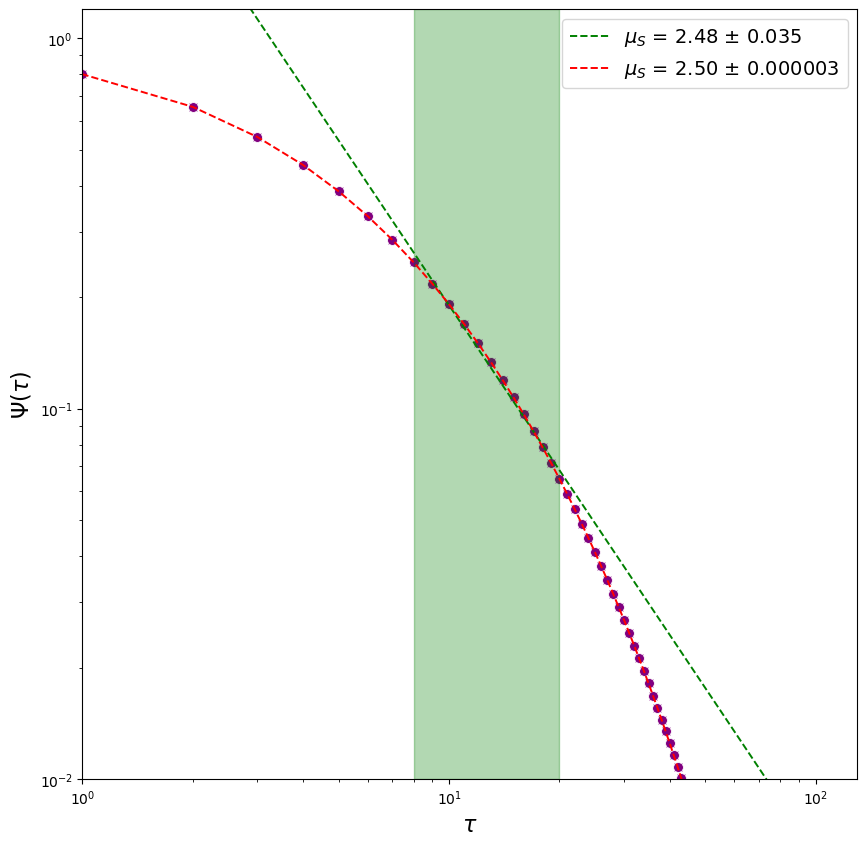}
\caption{ The plot shows the truncated survival probability generated using Eq. \ref{theoreticalsurvival}, the value of T = 7.8 and the value of $\Gamma = 0.04166 s^{-1}$ which mimics the survival probability of Figure 4.  The analytical fit is $\left(\frac{7.8}{7.8 + t}\right)^{\mu_S -1} e^{-\Gamma t}$ for the entire range of t values. The shaded region shows the range used for the straight line linear fit. }
\label{fig:9 }
\end{figure}

Having in mind the survival probability of Figure \ref{fig:4 }, we generated a sequence of times $\tau$ using the algorithm of Eq. (\ref{tau}). We set $T = 7.8$ to generate with this time series a survival probability as close as possible to the suggestions afforded by Figure \ref{fig:4 }, that the transition to the inverse power law begins after a time of that order. We select the value $2.5$ for $\mu$, which is denoted as $\mu_S$ to distinguish it from the inverse power law index $\mu_R$ of this Section. We assign to each laminar region $\tau$ either the value $1$ or the value $-1$, tossing a coin. This algorithm is expected to generate an exponential truncation at a time much larger than the time $1/\Gamma = 24 s^{-1}$ generated by Figure \ref{fig:4 }. This is so because this truncation would not have a physical origin but a merely algorithmic origin, due to the aging-induced ergodicity breakdown \cite{0}. For this reason the numerical survival probability generated by this algorithm is multiplied by $exp(-\Gamma t)$ with $\Gamma = 0.04166 s^{-1}$. The result is illustrated in 
 Figure \ref{fig:9 }. The similarity between Figure \ref{fig:9 } and Figure \ref{fig:4 } is remarkable.

 It is important to notice that the survival probabilities of Figure \ref{fig:4 } and Figure \ref{fig:9 } are also very similar to Figure 2 of \cite{3} showing that our theoretical arguments on the cancerous cells  generate dynamical properties equivalent to the intracellular transport of cargoes, a similarity that will guide our future work on the interaction between the cancerous cells. 

 This satisfactory similarity requires further research work to settle another open issue. In fact, we notice the time region before 
 the emergence of the inverse power index $\mu_S - 1 = 1.5$ has a smaller slope affording information on the time interval necessary to reveal the true complexity of the process. According to the theoretical and numerical arguments of \cite{1} the slower slope of this region is a manifestation of a stretched exponential, usually referred to as Mittag-Leffler (ML) function \cite{klafterml}. This waiting distribution density can derived using a proper extension of the L\'{e}vy central limit theorem \cite{pensri}. We note that the ML function involves $\mu < 2$, in conflict with the results of our statistical analysis yielding $\mu_S = 2.5$. We plan to devote future research work to discuss the extension of the ML arguments using the results of \cite{bologna20}.

\section{Regression Times} \label{section5}

We devote this section to the detection of another inverse power law index, $\mu_R$, which, in the case of a one-dimensional diffusion along the x axis refers to the time distance between two consecutive re-crossings of the same coordinate.

The theoretical and numerical work of this Section requires additional
discussion on the distinction between the two IPL indices $\mu_S$ and $\mu_R$ raised
by Failla et al. \cite{kpz}. The random growth of surfaces is a renormalization group
phenomenon \cite{recent} that may be interpreted as a process of self-organization.
In fact, the falling particles sticking to the side of the tallest nearby column establishes a cooperation process that has the effect of controlling the roughness of the growing surface. We explore this possibility herein with caution since it may have the effect of extending the criticality-induced intelligence to conditions
where life is not yet present. Here we limit ourselves to noticing that the connection between $\mu_S$ and $\mu_R$ was originally established for $\delta < 0.5$ \cite{kpz}. However, the result is valid under the condition $\delta < 1$. As a consequence it can be used also for the super-diffusion condition of the GB cells. The only significant change is on the evaluation of $\mu_R$. 

Herein we propose, for the emergence of intelligence, $\delta >0.5$. We cannot rule out the possibility that another form of intelligence
may emerge also for $\delta < 0.5$ as it may be hypothesized on the basis of the recent research of Kakin et al. \cite{recent} and Ikegami et al. \cite{criticalityandintelligence}, which would lead us
to consider also the condition $\delta < 0.5$ and $\mu_S <$  2,  consistent
with the theoretical approach of Haldar and Basu to KPZ
\cite{recent2}. 

We notice that also the authors of \cite{4} for the searching process proposed $\mu = 1.7$ corresponding to $\delta < 0.5$ and in the concluding remarks we shall make a plan to explore the motility of cancerous cells in search of nutrients that might lead them to adopt the sub-diffusional condition.  

In conclusion, some authors support the choice of $\delta < 0.5$ \cite{criticalityandintelligence, recent, 4}. For the reader's convenience we explicitly write Eq. (\ref{Exact}) as a function of $\mu_S$
in the sub-diffusional case
\begin{equation} \label{sub}
    \mu_R = 2 - \frac{\mu_S - 1}{2} = \frac{5 - \mu_S}{2},
\end{equation}
and in the super-diffusional case
\begin{equation} \label{sup}
    \mu_R = 2 - \frac{1}{\mu_S -1}.
\end{equation}

It is interesting to notice that using $\delta < 0.5$, $\mu_R$ as a function of $\mu_S$ changes from $\mu_R = 2$ for $\mu_S = 1$ to $\mu_R = 1.333$ for $\mu_S = 2$. The adoption of $\delta > 0.5$ yields $\mu_R  = 1$ for $\mu_S = 2$ and $\mu_R = 1.5$ for $\mu_S = 3$.

The adoption of $\delta > 0.5$ is compatible with reaching the condition of eternal survival, $\mu_R = 1$, whereas $\delta < 0.5$ gets the largest stability value, $\mu_R = 1.3333$ for $\mu_S = 2$.

\begin{figure}[H]
\centering
\includegraphics[width=0.6\textwidth]{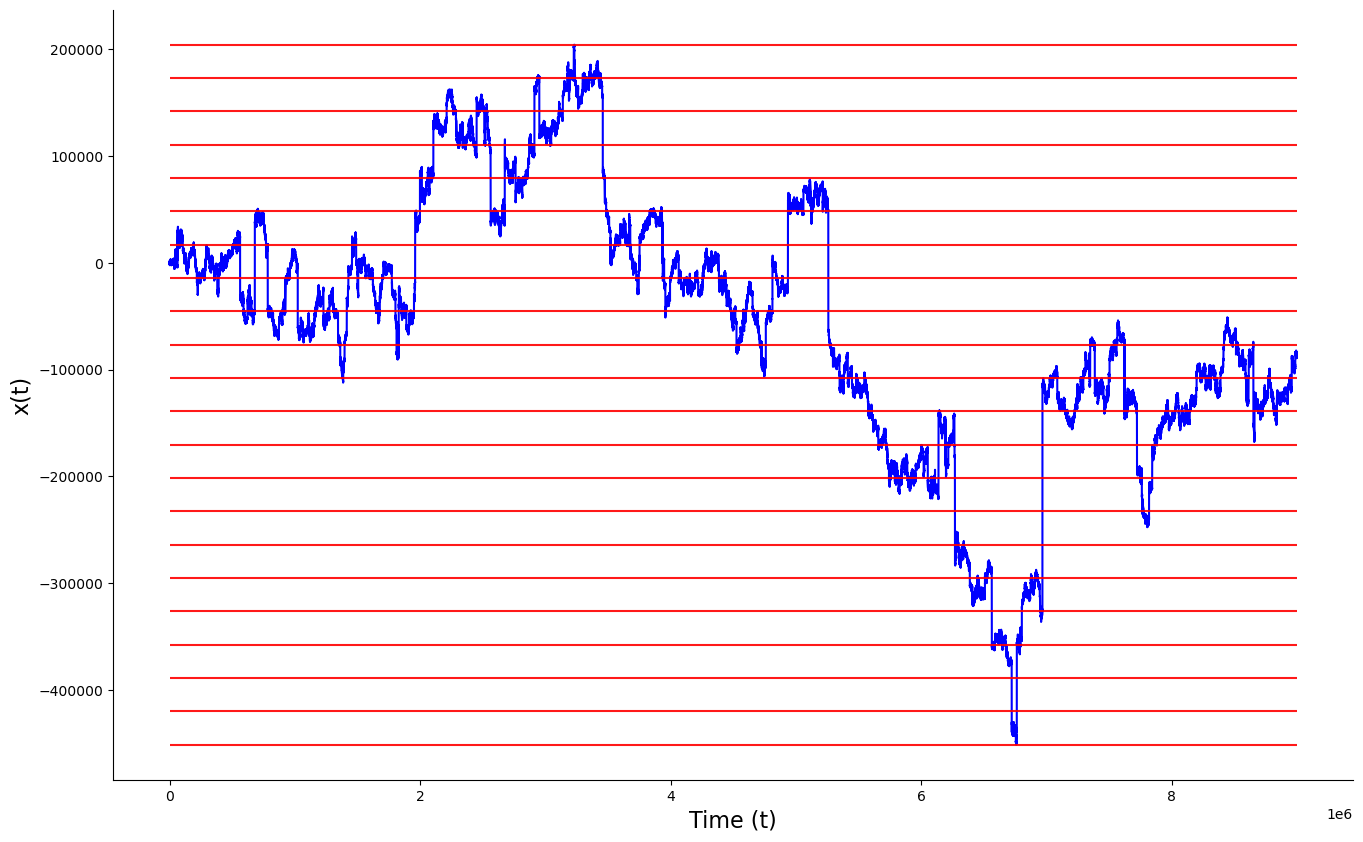}
\caption{Lévy walk trajectory with overlaid horizontal stripes used to obtain the times of recrossing. Each time the trajectory crosses a horizontal line (stripe), we recover an event corresponding to the recrossing of that level.}

\label{fig:10 }
\end{figure}

To study regression process for the L\'evy walk, let us study first the 1D case where the laminar regions are filled
with either +1 or -1, with equal probability. Figure \ref{fig:10 } shows the single trajectory $ x(t)$ made by a single random walker in 1D. The evaluation of $\mu_R$ through the analysis of the regression to the origin would require the use of extremely long trajectories. We made the convenient assumption that information on the regression to a level $x$, even when $x \neq 0$, contributes to the evaluation of $\mu_R$. The stripes
indicate that we extract the recrossing events for given thresholds, and they serve the purpose of detecting $\mu_R$ corresponding to the IPL waiting-time PDF for the time
intervals between consecutive re-crossings of the same level.

 It is very interesting to notice that our waiting times distribution density generating crucial events, according to \cite{crucialevents} is derived from the time dependent failure rate:

\begin{equation}
g(t) =  \frac{\psi(t)}{\Psi(t)} = \frac{r_0}{1 + r_1 t},
\end{equation}
which is identical to the proposal done by the authors of \cite{1} to generate super-diffusion rather than sub-diffusion. This is an interesting problem that may become the topic of our future research work.

\begin{figure}[H]
\centering
\includegraphics[width=0.5\textwidth]{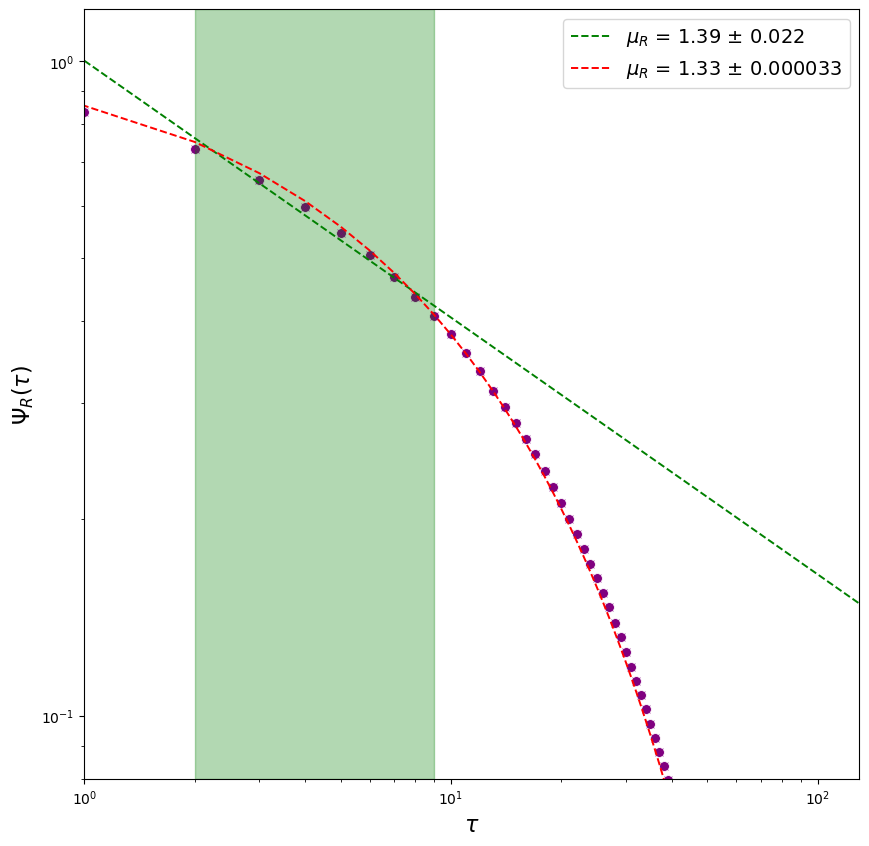}
\caption{Survival probability corresponding to the recrossings of Fig. \ref{fig:10 }, is fitted using two methods. The green line denotes a straight-line fit, and the red line is $\left(\frac{2.4}{2.4+t}\right)^{\mu_R -1} exp{-\Gamma t}$, where $\Gamma = 0.04166 s^{-1}$ and $ T_R $ used for the fit = 2.4. }

\label{fig:11 }
\end{figure}

 Thus we proceed with our approach to generate the inverse power index $\mu_R$, which is expected to be $\mu_R = 2 - 0.67 = 1.33$ on the basis of Eq. (\ref{Exact}).  To do that we use 
 the above described time series of $\tau$, filled with $\pm 1$ to generate the x trajectories illustrated by Fig. \ref{fig:10 }. We create many stripes with the same size. Following the prescription of \cite{kpz} we should establish the distance between two consecutive re-crossings of the origin. To get an accurate result without using the extremely large times we should use a number of re-crossings large enough to generate satisfactorily accurate statistics, we monitor the time distance between two consecutive re-crossings of the same (red) horizontal line in Figure \ref{fig:10 } , to define the waiting time distribution density and the more accurate survival probability. The resulting survival probability is multiplied by $e^{-\Gamma \tau}$ and the result is illustrated by Figure \ref{fig:11 }. 

 The result of Figure \ref{fig:11 } confirming the prediction $\mu_R = 2 - 0.67 = 1.33$ deserves a comment that may attract the attention of researchers with interest on the biologic importance of L\'{e}vy walks. We notice that the scaling of $\mu_S = 2.5$, $\delta = 1/(\mu_S -1)) = 0.666$ is identical to the scaling of $\mu_S = 1.666$ of the original KPZ effect \cite{kpz}, leading to $\mu_R = \mu_S = 1.666$. 

 This is the reason why we called $\mu_S = 2.5$ an ideal value. It is ideal because adopting the rule of making always a step ahead, not discussed in \cite{0}, it converts the KPZ scaling $\delta = 0.33333$ into the scaling $\delta = 0.66666$ emerging when the number of units of an organized network is $N = 150$ \cite{dunbar1,dunbar2}. Adopting for the motility of cancer cells the value $\mu_S = 2.5$ makes it possible to assign to the scaling of the diffusion process generated by the motility of the cancer cell the ideal value $\delta = 0.6666$.
 
 On the basis of this result, we are inclined to interpret the arguments of the authors of \cite{4}, with their $\mu =1.7$ as a tendency to select the criticality of KPZ, whereas the cancerous cell under study herein selects the same scaling making their internal organization less vulnerable. In fact $\mu_R = 1$ is a condition of eternal life and $\mu_R = 1.33$ is closer to $\mu_R = 1$ than $\mu_R = 1.66$. We believe that the cancerous cells studied in this paper make a balance between the two benefits, and co-opt for decreasing vulnerability.

For clarity we note that the survival probability for consecutive re-crossings of the same (red) horizontal line is expected to have the analytical form:

\begin{equation}
    \Psi_{R}(\tau) = \left(\frac{T_R}{T_R + \tau}\right)^{\mu_R -1} e^{-\Gamma \tau}, 
\end{equation}
which is used by us as an analytical formula to fit the numerical data illustrated in Fig. \ref{fig:11 }.

The result of Fig. \ref{fig:11 } confirms the validity of the relation between
$\mu_R$ and the scaling index given by Eq. (15). This result, going beyond the KPZ arguments of [5] leads us to the conclusion that Lévy
walk theory being used here reveals an important connection with criticality. 

\section{Lévy walk and criticality} \label{section6}

The work of Vicsek et al. \cite{vicsek} is a popular example of self-organization
frequently quoted to explain swarm intelligence. In the original paper the authors do not explicitly discuss the scaling index $\delta$ thereby
preventing us from stating that this popular model of phase transition fits the
central theoretical result obtained herein, namely, Eq. (\ref{Exact}). For this reason
we decided to evaluate $\delta$ and $\mu_R$ using their model.

\begin{figure}[H]
\centering
\includegraphics[width=0.6\textwidth]{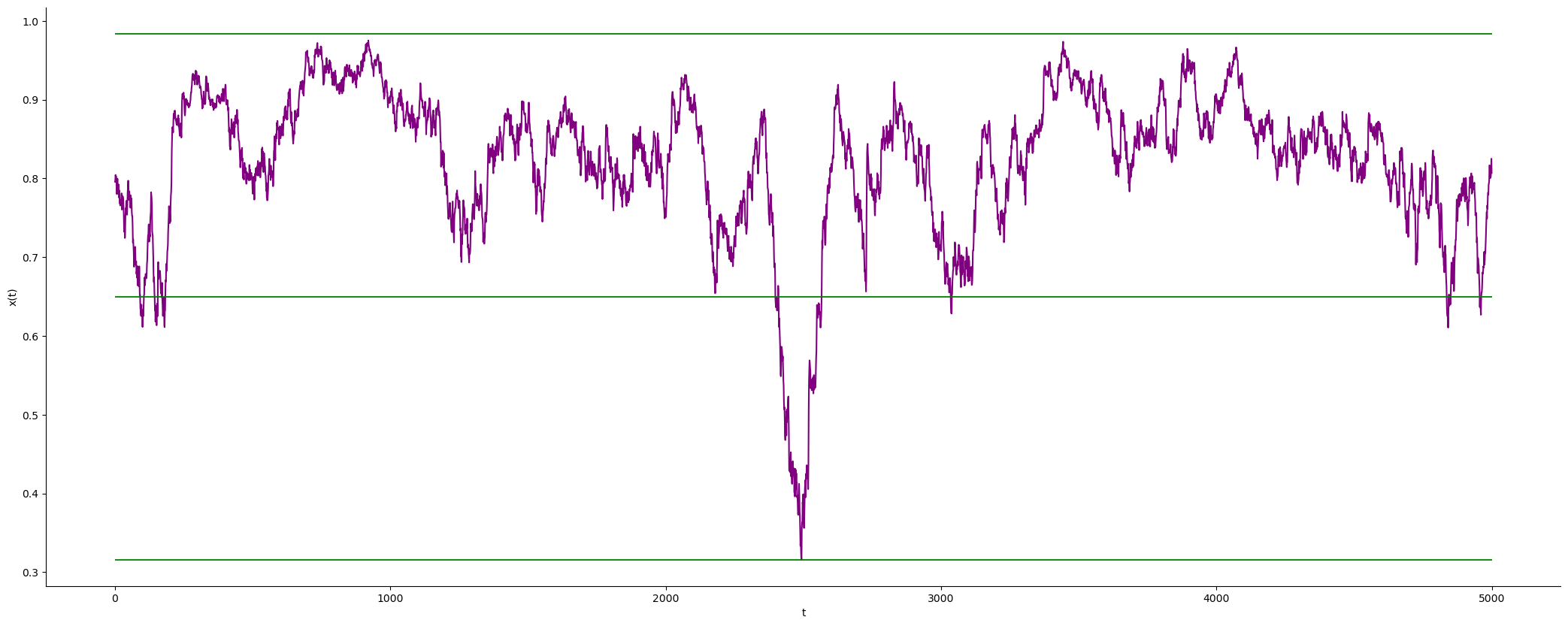}
\caption{We show the result of applying thresholds on the field x(t) generated from the Vicsek model \cite{vicsek}  with domain size = 32, noise =  0.6, particle number = 100, interacting radius = 1, absolute velocity = 0.5, and with periodic boundary condition. }
\label{fig:12}
\end{figure}

We generated the field $ x(t)$ plotted in Figure \ref{fig:12} with only two stripes to evaluate the value of $\mu_R$. The sequence of $x(t)$ is sufficiently large to evaluate the distribution of time distances between two consecutive recrossing of the intermediate line of Figure \ref{fig:12}. As far as the evaluation of the scaling $\delta$ is concerned in this case, we have to adopt a procedure different from that used to evaluate the scaling generated by the motility of the cancer cells, leading to the results of Figure \ref{fig:3 } and to $\delta = 0.67$. In this case the crucial events are invisible. Therefore we have to use the method of stripes proposed by \cite{dea3}. The adoption of this method leads us to the value $\delta = 0.666$, a remarkably encouraging result, since, according to Eq. (\ref{sup}) the value of $\mu_R$ should be 1.333. This result shows that the proposal made by the authors of \cite{dea3} to adopt the stepping ahead prescription is wise advice, leading to the same result as the velocity model in the region  $2 < \mu_S < 3$. In this case the CE-s are invisible preventing us from adopting the velocity model. The use of the stripes of \cite{dea3} generates events, either crucial or non-crucial. However, under the assumption that the non-crucial events yield the ordinary scaling
$\delta = 0.5$, in the long-time limit DEA perceives only the events generating $\delta > 0.5$, the crucial events of the condition $2 < \mu_S < 3$. We note that this is an assumption suggested by the results of cancer motility, where the CE-s are visible. However, the value of $\mu_R$ affords an indirect information on scaling. The adoption of $\delta = 2 -\mu_R$ recovering the same value for $\delta$ confirms the prediction of DEA with stripes \cite{dea3}, supporting at the same time our claim that $\mu_R = 1.333$ is an ideal value.

\begin{figure}[H]
\centering
\includegraphics[width=0.6\textwidth]{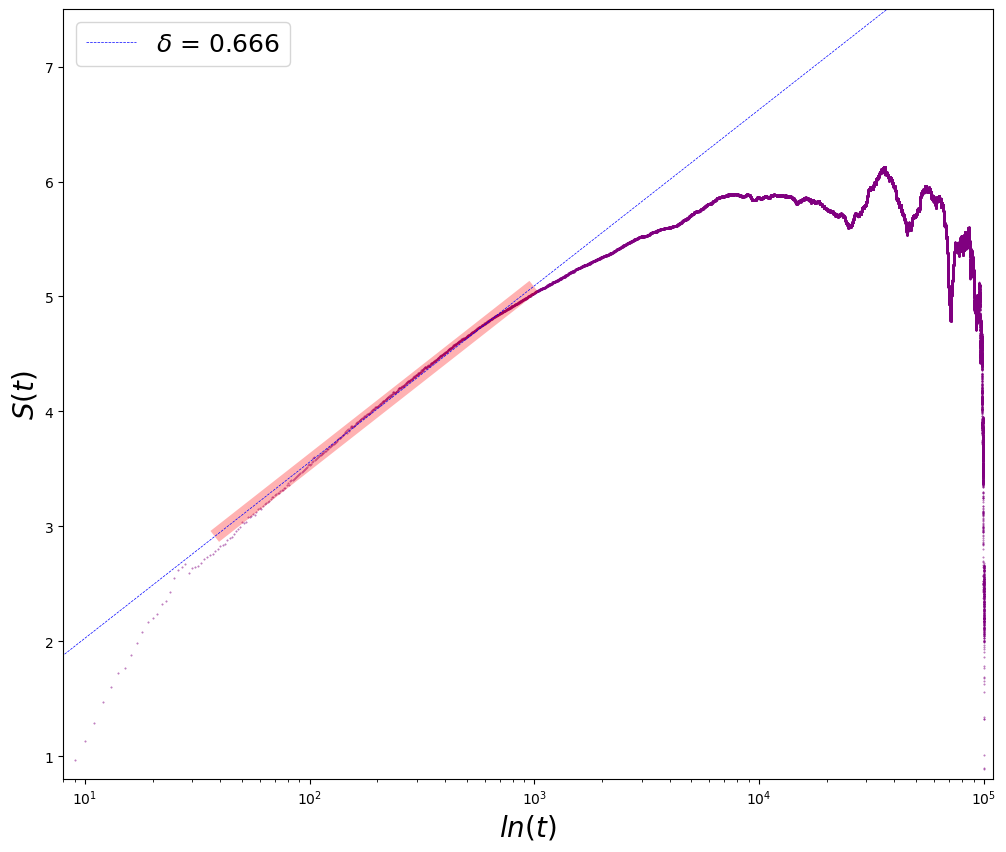}
\caption{The figure shows the results afforded by the method of DEA \cite{dea3} (with 100 stripes) for the scaling generated by the collective intelligence of \cite{vicsek}. The resulting value of $\delta$ is 0.666. }
\label{fig:13}
\end{figure}

Figure \ref{fig:13} depicts the entropy versus the log of time by applying DEA with stripes \cite{dea3}
to the data-set $ x(t) $ from Figure \ref{fig:12}, using one hundred stripes. The dashed line segment is fitted to the solid curve
yielding to $\delta$ = 0.666.

\begin{figure}[H]
\centering
\includegraphics[width=0.6\textwidth]{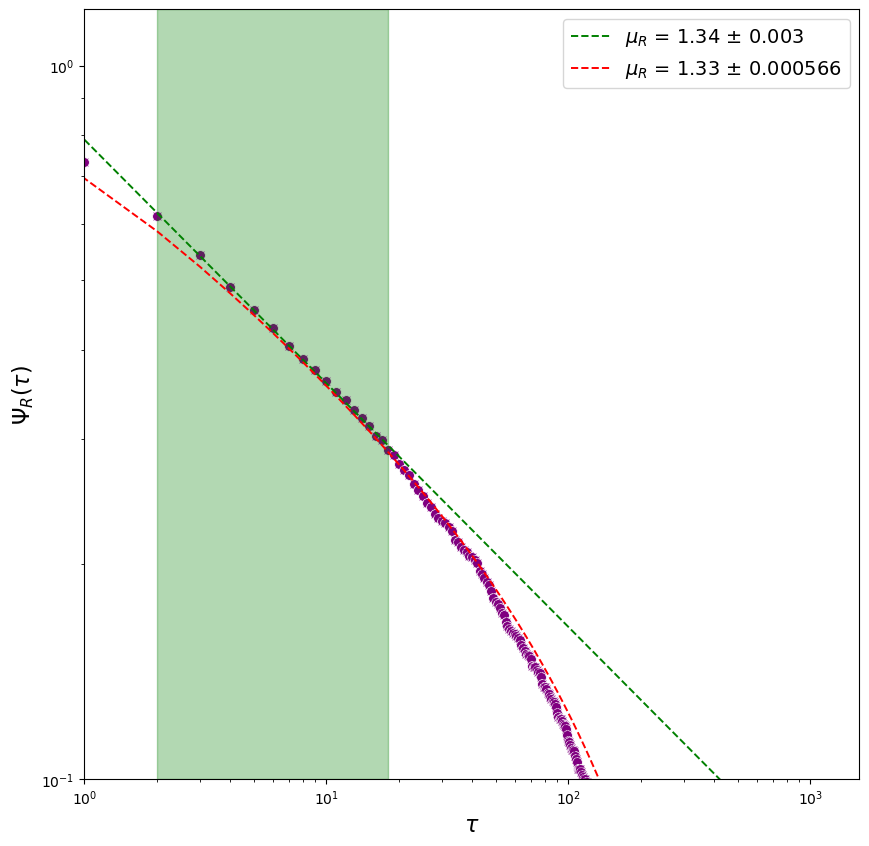}
\caption{Shown here are the results for $\mu_{R}$ obtained for x(t) of \cite{vicsek}, using the slope of the Survival probability $\Psi_R(t)$. The value of $\mu_R$ = 1.33, with the standard error of fit = $\pm$ 0.00056664. The survival probability is found to be truncated with a $T_R = 0.5$ and a $\Gamma = 0.0035 s^{-1}$ .}
\label{fig:14}
\end{figure}

Figure \ref{fig:14} depicts the survival probability $\Psi_R(t)$, obtained using only 2 stripes. It is important to notice that in this case we are observing the data generated by the theoretical approach to the swarm intelligence interpreted as a new form of phase transition \cite{vicsek}. The approach adopted to generate Figure \ref{fig:14} is based on the adoption of a waiting-time distribution density without truncation. We had to artificially introduce truncation to establish a closer connection with the experimental data of C. M. Mazzanti's lab. This suggests that truncation is a real property of criticality. The resulting  value $\mu_R = $
1.33, is in satisfactory agreement with the prediction of Eq.(\ref{Exact}). Truncation makes more difficult to find $\mu_R$ leading us to adopt the fitting procedure used for both Figure \ref{fig:9 } and Figure \ref{fig:11 }, but the result of this fitting procedure  confirms the importance of Eq.(\ref{Exact}), which is an
exact prediction based on the key assumption that the dynamics of criticality
are renewal.

It is of general interest that the discovery of $\mu_R = 1.33$ for $\mu_S = 2.5$, close to the analysis of experimental data on the motility of cancer cells, with $\delta = 0.67$ affords information on the vulnerability of cancer cells.

We note that according to \cite{mirkofabio}, the distance between two consecutive organizational collapses is illustrated by the survival probability: 

\begin{equation} \label{muRSurvival}
\Psi_R(t) = \frac{A^{\mu_R-1}}{(t+ A)^{\mu_R-1}},
\end{equation}
suggesting that the condition of maximal intelligence, $\mu_S = 2$, with scaling $\delta = 1$, makes a complex system invulnerable. 

\section{Concluding Remarks} \label{section7}

This paper illustrates an approach to measure the intelligence of the
single cancer cells and not the intelligence of a colony of GB
cells. According to Watson and Levin \cite{collectiveintelligence1} the intelligence of a complex
system is determined by the interaction between its sub-units, without ruling out the intelligence of the single cells. The
analysis of data done in this article focuses on
the emergent intelligence of a single cancer cell, since it is based on the
assumption that all the cells in the snapshot shown in Figure \ref{fig:2 } adopt the
same motion prescription, thereby allowing the use of the Gibbs perspective
of an ensemble.
We conclude that a single GB cell has two distinct
IPL indices $\mu_S = $  2.5 and $\mu_R =$ 1.33 that we interpret to be measures of the self-organization of the components of a single cell and of its breakdown, generating the important question
of whether a single cell may have a criticality-induced ``intelligence". We proffer that
this conclusion is supported by evidence therefore upholding the conjecture that a
single GB cell has a criticality-induced ``intelligence". This assumption
is not in conflict with the opinion of professional biologists, see e.g. the work
\cite{collectiveintelligence1}. The decision making capability of single cells seems to be confirmed by
the even more recent work of \cite{singlecell}.
The important role of criticality is frequently adopted. However, the
fact that criticality, and self-organized criticality as well, generate CE-s is not yet adequately taken into account. The research work currently done on the mobility of cells is of increasing interest \cite{ motility2, motility3, motility4} and
the purpose of this paper is to attract the attention of the investigators in this
field of research on the connection with both the emergence of intelligence and organizational collapse. The
experimental results illustrated by Figure \ref{fig:2 } suggest that criticality generate the CE-s as abrupt changes of swimming direction, making the dynamics of the
cells similar to the flying of albatross, widely interpreted as a manifestation
of intelligence. This has the effect of shedding light into the research work
on physiological processes, like the dynamics of the human brain and the
heartbeats. It is of remarkable interest that the motion of the cells projected
along the $x$ axis and the $y$ axis, with the CE-s becoming invisible,
yields the same anomalous scaling of physiological processes.
Herein, in accordance with \cite{singlecell} we find that Lévy walk is a manifestation of intelligence of a single cell. We make the conjecture that $\mu_S = 2.5$ is a manifestation of CE-s interpreted according to \cite{dunbar2} as failures 
instantaneously repaired by the cooperation between the units of the single
cell. 

There is probably a balance between the tendency to select $\mu_S = 1.7$ that may favor the searching process \cite{4} and $\mu_S = 2.5$ generating the same scaling as $\mu_S = 1.6666$ when DEA is applied using the always jumping ahead prescription \cite{dunbar1, dunbar2}. The cancerous cells may get the benefit of lower vulnerability adopting $\delta > 0.5$ rather than $\delta <0.5$ leading to the traditional KPZ effect. 

The discovery of the connection between criticality-induced intelligence
is signaled by the important relation $\mu_R = 2 - \delta$ that at the same time may
afford important suggestions to make the scaling of cancer cells depart as much as possible from the scaling $\delta =1$ making them invulnerable. 

We plan to investigate the possibility of increasing vulnerability of cancerous cells by stimuli activating searching activity as a method to control cancer. The integrin $\alpha5\beta1$ facilitation of cancer cell invasion \cite{5} may at the same time make them more vulnerable. 
The lab C.M. Mazzanti is currently using nutrients to activate the searching process of the GB cells, and we plan to study the dynamics of the cancerous under the influence of these stimuli.
We noticed that the authors of \cite{2}, not focusing on cancerous cells, proposed a theory leading to $\mu_S = 2$, which would correspond to $\mu_R = 1$, the invulnerable condition of Eq. (\ref{muRSurvival}) . We have not yet analyzed the dynamical behavior of healthy cells. This will be the subject of future research work, where the distinction between $\mu_R = 1.33$ for cancerous cells and $\mu_R = 1$ for healthy cells would be of crucial importance for the control of GB. 

As the last but not least issue, we stress the connection between the criticality-intelligence issue of this manuscript and the L\'{e}vy foraging patterns of rural humans \cite{4} allowing this paper to
transition from biological complexity to the field of social complexity, a topic of increasing interest. 

A limitation of the present paper has to do with the fact that the current experimental results 
of \cite{mazzanti} do not allow us to establish if the intelligent dynamics of the single cell is limited to the criticality generating the decision making process of a single cell \cite{singlecell} or is already influenced by the cell-to-cell communication. This will be the topic of further research work. 

The shortness of the experimental trajectories we analyze prevents us from 
computing the scaling of the diffusion of each different cell. This should be
an important goal to be achieved but it would need very long experimental videos
that probably lie beyond the biological experimental constraints. 

Nevertheless, the future availability of new and longer data can lead to a classification
of cells in different categories according to their mobility. In this case, we would proceed
to evaluate the scaling of the most dynamics, medium, and lazy cells.  
This approach can be important because it is probable that the spreading of cancer
cells in the organism is driven by the most dynamic cells inside a cancer tissue and
not by the mean mobility of the cells.

\emph{Acknowledgments} : We extend our heartfelt gratitude to Dr. Anna Luisa Di Stefano for their invaluable contributions to patients’ sample collection and clinical management, and for their scientific advice. Additionally, we would like to express our sincere appreciation to Dr. Lessi and Dr. Sara Franceschi for their essential roles in processing brain samples and providing valuable scientific guidance throughout the project's development. We would like to express our deep gratitude to an unknown reviewer who made comments that helped us to significantly improve the paper.

\end{document}